\documentstyle[12pt]{article}

\begin{document}
\def\ii{\'{\i}}
\def\bi{\bigskip}
\def\be{\begin{equation}}
\def\en{\end{equation}}
\def\bq{\begin{eqnarray}}
\def\eq{\end{eqnarray}}
\def\noi{\noindent}
\begin{center}
{\Large \bf CENTRAL CHARGES IN REGULAR MECHANICS\footnote{Work supported by CONACyT under contract 3979PE-9608}}\\[1.5cm]
\end{center}
\begin{center}
{{\large \bf A. Cabo\footnote{On leave of absence from ICIMAF, La Habana, Cuba}, J.L. Lucio M. and V. Villanueva}\\[.3cm]
{\it Instituto de F\'\i sica, Universidad de Guanajuato}\\
{\it Apartado Postal E-143, Le\'on, Gto., M\'exico}}
\end{center}

\vspace{1.8cm}

\begin{abstract}
\setlength{\baselineskip}{.5cm}
We consider the algebra associated to a group of transformations which are
symmetries of a regular mechanical system ({\it i.e.} system free of 
constraints). For time dependent coordinate transformations we show that a
central extension may appear at the classical level which is coordinate and 
momentum independent. A cochain formalism naturally arises in the argument  
and extends the usual configuration space cochain concepts to phase space.
\end{abstract}

\setlength{\baselineskip}{1\baselineskip}

\newpage

\section{Introduction}

\bi
\bi

The concepts of cochains and cocycles have demonstrated to be of relevance for
the discussion of anomalous behavior of symmetries in QFT [1]. The
applicability of these concepts has spread to classical and quantum mechanics 
providing a mathematical framework in which symmetry and symmetry breaking can
be analyzed [2].

\bi

In a previous work [3], we have proposed a way to obtain a quantum mechanical
geometrical phase for a classical system which has the characteristic that the
action, but not the Lagrangian, is invariant ({\it i.e.} when the variation of 
the Lagrangian is a total time derivative) under contact coordinate 
transformation. We have encountered difficulties in applying this formalism
to problems such as SUSY quantum mechanics [4] or scale invariance in two 
dimensional quantum mechanics [5]. The problem is that the cochain structure 
appearing in these problems are velocity dependent while the conventional
approach to cochains is done in configuration space. The aim of this paper
is two fold. First, we discuss the variation of the phase space Lagrangian
({i.e.} the Lagrangian written in terms of canonical variables) under 
{\bf finite} transformations. This enable us to apply the cochain formalism 
in phase space. The second objective is to consider physical systems
possessing a group of symmetry which is considered in order to analyze the 
possibility that the Poisson brackets of Noether's charges acquire a central
extension.

\bi

Noether's theorem provides a systematic way of analyzing the conserved 
quantities associated to a physical system. The conventional approach consists 
in showing the invariance of the action under transformations for\-ming a 
continuous group $G$ of dimension $\omega$. Noether's theorem then assure the
existence of $\omega$ conserved charges $Q_r$. In the Hamiltonian formalism,
the Poisson brackets of the conserved (Noether's) charges define an algebra
which is isomorphic to the algebra of the global symmetry group from
which the charges were obtained [6]. As a consequence, the charges $Q_r$
generate, through their Poisson brackets, the corresponding global symmetry
transformations of phase space variables. There exist however the possibility 
that, the algebra of the charges $Q_r$ is an extension of that of the global 
symmetry group. At the quantum level the same statement applies to the 
commutators of the charges, however due to ordering of composite operators, 
new terms that vanish in the classical limit $(\hbar \to 0$) may appear in the
associated commutation relations, indicating the existence of an anomaly, 
{\it i.e.} the breaking of the classical symmetry by quantum effects [7]. A 
similar phenomenon may occur at the classical level, indeed as a consequence 
from the passage from configuration to phase space, new terms -as compared to 
the algebra of the original group of transformation $G$- may appear in the 
Poisson brackets of the $Q_r$ charges. According to our results, a necessary 
condition for this to happen is that the action but not the Lagrangian be
invariant under the symmetry transformation. The Galilei group and the 
magnetic translation group provide examples where such conditions are met
and a classical central extensions appear.

\bi

We tried to make the text as self contained as possible, to this end we have 
included in section 2 a short summary of Noether's theorem and the 
corresponding expression in phase space. Section 3 and 4 are devoted to the
analysis of a possible central extension of the algebra of Noether's charges.
In particular, in section 3 we consider coordinate transformations for which
the variation of the Lagrangian in phase space leads to a cochain structure 
that implies the existence of conserved quantities. We show that such a
constant of motion is related to the central extension. Section 4 is devoted 
to generalize the previous results, in this case however, the cochain structure
does not enter the derivation of the central extension.

\bi

In order to see the ideas underlying our approach in a concrete setting, we
consider the following three physical systems: {\it i)} Motion of a
particle in two dimensions under the influence of a scale invariant potential.
In this case, both the Lagrangian and the action are invariant under the 
transformations and there is no central extension of the algebra of Noether's
charges. {\it ii)} Group of magnetic translations. This system concerns the
movement in two dimensions of a charged particle in a homogeneous magnetic 
field. The symmetry transformations to consider are translations. The central
point is the incorporation of the vector potential, which lead both to the
non-invariance of the Lagrangian and the modification of the translation
generator. The central charge is a consequence of the non-vanishing Poisson 
bracket of the momentum and the vector potential. {\it iii)} Galilei 
invariance. This is a well known example [8] where a central extension of the 
algebra appears. We work out details of the calculation and show the relation
between the central extension and the non invariance of the Lagrangian (due to
the surface term) under Galilei tranformations.

\bi
\bi

\section{Noether Theorem}

\bi

We consider a system with $n$ degrees of freedom, given as functions $q_j (t)$
$(j= 1,2,\cdots n)$ of the ``time" variable $t$. We assume that the dynamics   
of the system is described by the action functional: 

\be
S \left[ q_j \right] = \int^{t_f}_{t_i} dt {\cal L} (q_j, \dot q_j ). 
\en

\noi Here, ${\cal L} (q_j, \dot q_j)$ is the Lagrange function depending on the
generalized coordinates $q_j$ and their corresponding velocities $\dot q_j$, 
but not on time. The time evolution of the system is described by a set of 
~$n$~ second order differential equations (the Euler-Lagrange equations of 
motion), which are linearly independent if the Hessian of the Lagrangian has a 
non-vanishing determinant,

\[ 
det \Big( \frac{\partial^2 {\cal L}}{\partial \dot q_i \partial \dot q_j} \Big)
\not = 0  .
\]

\noi When this condition is met, the system is said to be regular. In 
the opposite case, when the Hessian has zero modes, the system is said to be
singular, and is characterized by the existence of constraints. In this paper 
we restrict our selves to the study of regular systems.

\bi

Let us consider transformations of the form:

\bq
q_j (t) &\to& q^\prime_j (t^\prime) = f_j (q(t), t, \alpha ) \nonumber \\
t &\to& t^\prime  = f_0 (t,\alpha) . 
\eq

\bi

\noi In Eq. (2) $q_i(t)$ are the coordinates in a time slice $t$ in 
configuration space and $q^\prime_i (t^\prime)$ is the image point of $q_i(t)$
at the time slice $t^\prime$, $\alpha$ stands for the set of $w$ parameters 
specifying the transformations and $q(t)$ is used to denote collectively the 
$n$ coordinates. The parametrization is chosen in such a way that:

$$f_j (q(t), t,0) = q_j (t), \quad f_0 (t,0)=t \eqno (3a)$$ 

\noi and

\[ f_j (f(q(t), t,\alpha), f_0 (t,\alpha),-\alpha) = q_j (t) \]
$$f_0 (f_0 (t, \alpha), - \alpha) = t , \eqno (3b)$$

\bi

\noi that is, for $\alpha =0$ Eq. (2) reduces to the identity and the 
inverse transformation is obtained by reversing the sign of the $\alpha$ 
parameters.

\bi

We assume that these transformations define a continuous group $G$ of dimension
$w$. Before establishing the relation of Eq. (2) with Noether's charges, we 
introduce the structure constants associated to $G$. For infinitesimal 
transformations with parameters $\delta \alpha_r$ we write $(\delta \alpha_r =
\frac{\alpha_r}{N}$ with N arbitrary large, and unless otherwise stated, here 
and thereafter sum over repeated indices is assumed):

\setcounter{equation}{3} 

\bq
q^\prime_j (t^\prime) &=& q_j (t) + \delta q_j = q_j (t) + (\delta \alpha_r 
T_r) q_j , \nonumber \\
t^\prime &=& t+ \delta t = t + (\delta\alpha_r S_r) t ,
\eq

\noi where $T_r, S_r, \,\, r= 1,2, \cdots ,$ $\omega$ are the group 
``generators" (appropriated algebraic or differential operators). 
A comment about Eq. (4) is necessary. Notice that Eq. (2) implies that
the coordinate and the time transformations depend upon the same set of 
parameters $(\alpha_1 , \alpha_2 ,\ldots ,\alpha_\omega)$. Furthermore, 
the time transformations depends only on time and the $\alpha_r$ 
parameters, therefore such a set of transformations must form a group  by
itself. That is the reason to include $\omega$ generators $S_r$ in (4). It
may happen however that the time transformations involve only some of the 
$\delta \alpha_r$. The following examples may be useful in clarifying these 
points.

\bi

\begin{itemize}
\item Consider scale transformation, defined by

\bq
q_i &\to& q^{\prime}_i = \frac{1}{\sqrt{1+\alpha_1}} q \nonumber \\
t &\to& t^\prime = \frac{1}{(1+\alpha_1)} t .
\eq

If we perform a second transformation on the coordinates we obtain:
$q^{\prime\prime}=\frac{1}{\sqrt{1+\alpha_2}}\,\,\frac{1}{\sqrt{1+\alpha_1}}\,
q=\frac{1}{\sqrt{1+\alpha_1+ \alpha_2 + \alpha_1\alpha_2}} q$. Which is of the
type (5) and consequently the coordinate transformations form a group if the 
composition law $C(\alpha_1 ,\alpha_2)\equiv \alpha_1 +\alpha_2 +\alpha_1 
\alpha_2$ is assumed. The point to emphasize is that the same reasoning holds 
for the time transformation. 

\bi

For infinitesimal $\alpha_1$ we obtain

\bq
\delta_1 q_i &=& -\frac{\alpha_1}{2}q_i , \,\,\,\,\,\, (i=1,2) \nonumber \\
\delta_1 t &=& -\alpha_1 t \nonumber
\eq

\noi therefore, the corresponding coordinate ($T_i$) and time ($S_i$) 
generators are given by

\be
T_1 =- \frac{q_j}{2} \, \frac{\partial}{\partial q_j}, \qquad\qquad S_1=- t
\frac{\partial}{\partial t}
\en

\item As a second example we consider Galilei transformations:

\bq
q^\prime_1 &=& q_1 + v t+a, \,\,\, q^\prime_2= q_2, \nonumber \\
t^\prime &=& t+ b. \nonumber
\eq

\noi To simplify our discussion we take $a=b=0$. In this case, there is no 
variation of the time $\delta_2 t=0$, whereas $\delta_2 q_1 = vt$. The 
corresponding generators are

\[  T_2 =\frac{\partial}{\partial q_1} , \,\,\,\,\,\ S_2 =0 \]

\bi

A further step will be to consider both scale and Galilei transformation.
It is easy to check that the $T_i$ generators $(i=1,2)$ close and that the 
structure constants obtained are the same as those entering in the 
commutation relations for the $S_i (i=1,2)$ generators.
\end{itemize}

\bi

Coming back to our general discussion, finite transformations can be obtained 
in terms of the $T_r ,S_r$ generators by exponentiating (4) 

\bq
q_i^\prime (t^\prime) &=& e^{\alpha_rT_r} q_i (t) \equiv g_c (\alpha) q_i (t), \nonumber \\
&~ & \mbox {}\\
t^\prime &=& e^{\alpha_r S_r} t \equiv g_\tau (\alpha) t . \nonumber
\eq

The group property of the transformations (2), expressed either for $g_c$ or
$g_\tau$ as

\be
g (\alpha_i) g (\alpha_j) = g (c(\alpha_i, \alpha_j)) ,
\en

\noi can be used to bring out the algebra of the generators. This is 
achieved by considering the commutator of two infinitesimal transformations: 

\[ g (\alpha_r)g(\alpha_s)-g(\alpha_s) g(\alpha_r)= g(c(\alpha_r, \alpha_s))-
g (c(\alpha_s , \alpha_r)), \]

\bi

\noi the Taylor expansion of these expressions leads to:

\be
\left[ T_r, T_s \right] = C^u_{rs} T_u , \,\,\,\, \left[ S_r, S_s \right]
= C^u_{rs} S_u 
\en

\noi where the structure constants $C^u_{rs}$ are defined as:

\be
C^u_{rs} =\frac{\partial c^u (\alpha_r,\alpha_s)}{\partial \alpha_r \partial 
\alpha_s}\Big|_{\alpha_r =\alpha_s=0} - \frac{\partial c^u(\alpha_s,\alpha_r)}{\partial \alpha_r \partial \alpha_s} \Big|_{\alpha_r =\alpha_s =0}
\en 

\bi

For latter use, it is convenient to express this property in terms of the
time and coordinate variations

\bq
\delta q_{_j} &=& q^\prime_j (t^\prime)-q_j(t)=\delta\alpha_r \frac{\partial 
f_j(q, t,\alpha)}{\partial \alpha_r}\Big|_{\alpha =0} \equiv \delta\alpha_r
\delta^r q_{_j} \nonumber \\
\delta t &=& \delta\alpha_r \frac{\partial f_0 (t,\alpha)}{\partial \alpha_r}
\Big|_{\alpha =0} \equiv \delta\alpha_r \delta^r t . \nonumber
\eq

\noi Given the group property of the coordinate transformation

\[ q_j^{r,s}=f_j(f_j(q,t,\alpha_r), f_0(t,\alpha_r),\alpha_s)
=f_j (q,t,\alpha_{rs}) , \]

\noi we calculate

\[ q^{r,s}_j - q^{s,r}_j =f_j (q,t,\alpha_{rs})-f_j(q,t,\alpha_{sr}) ,\]

\noi which for infinitesimal transformations results in

$$q^{r,s}_i-q^{s,r}_i =\Big(\delta^r q_j \frac{\partial}{\partial q_j}\delta^s
q_i -\delta^s q_j\frac{\partial}{\partial q_j} \delta^r q_i + \eqno (10a)$$
\[ \delta^r t\frac{\partial}{\partial t} \delta^s q_i -\delta^s t\frac{\partial}{\partial t} \delta^r q_i \Big) = C^u_{rs} \delta^u q_i , \]

\bi 

>From Eqs. (10, 10a) we see that the structure constants $C^u_{rs}$ are defined
through the composition law of the group elements. On the other hand, we 
already pointed out beneath Eq. (4) that the coordinate and time transformation
must have the same composition rule, therefore the commutator among the
$S_r$ generators is $[S_r, S_s] = C^u_{rs} S_u$.

\bi

Let us consider now the relation between the transformation (2) and the 
physical system. At the classical level the system described by (1) is said to
possess a symmetry or to be invariant if, up to surface terms, the action is 
form invariant under the transformations (2). In terms of the Lagrangian, this 
property is equivalent to the requirement

\bi

\setcounter{equation}{10} 

\be
\Big( \frac{dt^\prime}{dt}\Big) {\cal L} \Big(q^\prime , \Big(\frac{dt}{dt^\prime}\Big) \frac{dq^\prime}{dt} \Big) = {\cal L} (q, \dot q)+ \frac{d\Lambda (q)}{dt} .
\en   

\bi

\noi When $\frac{d\Lambda}{dt}=0$ the Lagrangian is said to be invariant under
(4). For in\-fi\-ni\-te\-si\-mal variations and to first order, Eq. (11) 
reduces to the identity [6]:

\bi

\[ \frac{d}{dt}(\tilde Q_r\delta \alpha_r)=\sum_j (\delta q_j -\delta t \dot q)
\Big( \frac{\delta {\cal L}}{\delta q} - \frac{d}{dt} \frac{\delta {\cal L}}{\delta\dot q} \Big) , \] 

\bi

\noi where

\be
\tilde Q_r(q,\dot q,t)\delta\alpha_r=\frac{\partial{\cal L}}{\partial\dot q_j} 
\delta q_j -\Big( \frac{\partial {\cal L}}{\partial \dot q_j} \dot q_j- 
{\cal L} \Big) \delta t - \Lambda . 
\en 

\bi

This is Noether's theorem, which implies that for any classical solution to 
the equation of motion there are $w$ constants of motion, or conservation 
laws. 

\bi

In the Hamiltonian formalism, the charges $\tilde Q_r$ generate, through their
Poisson brackets, the global symmetry transformations on phase space. 
In order to analyze this property and possible generalizations, we are 
naturally lead to the study  of the conserved charges in phase space.
    
\bi

The Hamiltonian $H$ is given by

\be
H (q, p)= p_i\dot q_i - {\cal L} (q, \dot q) ,
\en 
\noi with
\be 
p_i = \frac{\partial {\cal L}}{\partial \dot q_i} , \qquad
\dot q_i = \frac{\partial H}{\partial p_i} .
\en 

\bi

\noi Since we are studying regular systems, then the conserved charges can be 
expressed in terms of the canonical variables 

\be 
\tilde Q_r (q, \dot q, t) = Q_r (q,p,t)= \tilde Q_r (q, 
\frac{\partial H}{\partial p} ,t) , 
\en 

\noi thus the charges take the phase space form:

\be
Q_r (q,p,t)\delta \alpha_r =p_i\delta q_i-{\cal H} (q,p) \delta t-\Lambda .
\en

Charge conservation in phase space is expressed as: 

\be     
0=\frac{d Q_r (q,p,t)}{dt} =\frac{\partial Q_r}{\partial q_\ell} \dot q_\ell + 
\frac{\partial Q_r}{\partial p_\ell} \dot p_\ell + \frac{\partial Q_r}{\partial t} .
\en

\noi Assuming that through any point  of  phase  space  can  pass a  solution
(there  are no constraints), it follows that charge conservation is expressed
in phase space in terms of the Poisson brackets:
 
\be 
\{ Q_r, H\} + \frac{\partial Q_r}{\partial t} = 0 . 
\en

\noi We will refer to the linear operator acting on $Q_r$ in Eq. (18) as the 
time Lie derivative. Associated to each infinitesimal transformation in 
configuration space $g(\delta\alpha_r)$, we have an infinitesimal canonical 
transformation 

\bq
q^g_i (q,p) &=& q_i-\delta \alpha^r \{ Q_r, q_i\} = q_i + ~\delta_c q_i , \\
p^g_i (q,p) &=& p_i-\delta \alpha^r \{ Q_r, p_i\} = p_i + \delta_c p_i .
\eq

\bi

\noi where the subindex $c$ indicates that these are increments due to 
canonical transformations generated by the $Q$ which are related to $\delta q$
of Eq. (16) by $\delta_c q=\delta q -\frac{\partial H}{\partial p} \delta t$. 
If we restrict our attention to classical configurations corresponding to 
solution to the equations of motion we have:   

\[ \{ q^g_i (q,p), H (q^g, p^g)\} =\delta\alpha^r \left[\{- \{ Q_r,q_i \}, 
H(q,p)\}+ \{ q_i,- \{Q_r, H (q, p)\}\}\right] + \frac{d q_i}{dt} \]
\be
= \frac{d q^g_i(q,p)}{dt}.
\en 

\noi In a similar way it follows that the  transformed momentum  $p^g (q, p)$  
sa\-tis\-fies the original canonical equation. Then, the mappings (19) and
(20) are sy\-mme\-tries of the Hamiltonian system and the charges $Q_r$, 
obtained from the Lagrangian  conserved  charges (12), generate symmetry 
transformations in phase space.

\bi

Finite canonical transformation are built in terms of the $Q_r$ charges by 
using the exponentiation of the generators in (19) and (20).

\be
{\cal U} (g(\alpha)) = e^{-\{\alpha^r Q_r,\,\,\,\} } ,
\en

\noi where the symbol $\{\alpha^r Q_r,~\}$ in the exponential, means a Poisson 
bracket understood as a linear operator acting on functions of the phase space
points. It proves convenient to parametrize the transformation (22) in terms of
a real arbitrary parameter $\sigma$ and a unit vector $(s_1, s_2 \ldots , s_w)$

\[ {\cal U}(g(\alpha))={\cal U}(s,\sigma)=e^{\{\sigma Q_r (q,p,t)s_r, ~\}}. \] 

\noi Thus, finite transformation of the coordinates and momenta are given by 

\bq
q^g_i (q,p) &=& {\cal U} (s,\sigma) q_i , \nonumber \\ 
p^g_i (q,p) &=& {\cal U} (s,\sigma) p_i .
\eq

\bi
\bi
\bi

\section{Phase Space Cochains and Central Char\-ges}

\bi
\bi

So far we have summarized infinitesimal symmetry transformations both in
configuration and phase space, including Noether's theorem and the 
asso\-cia\-ted conserved charges, which serve as generators of the 
transformations in the Hamiltonian formalism. In this section we prove that 
the variation of the Lagrangian under finite transformations is given by a 
time Lie derivative. This results allow us to introduce the cochain formalism 
in phase space and conclude that, under very specific conditions, a coordinate
and momentum independent central extension of the algebra arises.

\bi

We begin with a brief reminder of the cochain and coboundary concepts [1].
Consider a transformation $g$ which belongs to a group of transformations.
Suppose $g$ acts on abstract space variables according to a definite rule,

\[ x \overrightarrow{_{_{g}}} x^g \]

\noi and the group composition law is

\[ g_1 g_2 = g_{12} . \]

\bi

\noi The application of two successive transformation yields \footnote{Given a 
symmetry group of a classical system, this property holds in configuration 
space. However, once we go over phase space, this need not be the case.}

\bi

\[ x \overrightarrow{_{_{g_1}}} x^{g_1} \overrightarrow{_{_{g_2}}} 
(x^{g_1})^{g_2} = x^{g_{12}} . \]

\bi

Quantities ~that ~depend ~on ~$x$ ~and ~$n$ ~group ~elements ~are ~called 
~n-cochains $\omega_n (x, g_1, g_2,\ldots g_n)$. The coboundary operation 
$\Delta$ is defined as:

\bq
\Delta \omega_n &\equiv& \omega_n (x^{g_1}; g_2, \ldots , g_{n+1})-\omega_n
(x; g_{12}, g_3 ,\ldots , g_{n+1}) + (-)^m \nonumber \\
&& \omega_n (x; g_1, \ldots , g_{mm+1}, \ldots , g_{n+1})+ (-)^{n+1}
\omega_n (x; g_1,\ldots , g_2). 
\eq

\bi

\noi The coboundary has the important property that $\Delta^2 =0$. Further 
details about cochains, cocycles and coboundaries can be found in [1] and 
references there in.

\bi
 
The Lagrangian can be expressed in terms of canonical variables as follows

\bi

\be
{\cal L} (q,p)=p_i\frac{\partial {\cal H}}{\partial p_i}-{\cal H}(q,p). 
\en

\noi A finite transformation of the Lagrangian is obtained in the following
way:

\bq
{\cal L} (q^g, p^g) &=& \Big( {\cal U} \Big( s, \frac{\sigma}{N} \Big)\Big)^N {\cal L} (q,p) \nonumber \\
&=& \left[ e^{-\{\frac{\sigma}{N} s_r Q_r, ~\}} \right]^N {\cal L} (q,p) . 
\eq

\noi The finite transformation has been expressed as the product of a large 
number N of identical infinitesimal mappings 

\be
e^{-\frac{\sigma s_r}{N}\{Q_r, ~\}} {\cal L} (q,p) \cong {\cal L} (q,p) - 
\frac{\sigma s_r}{N} \{ Q_r, {\cal L} (q,p)\} + \ldots
\en

\noi Using the Jacobi identity and the conservation law (18), the Poisson 
bracket is rewritten as:

\bq
-\frac{\sigma s_r}{N} \{ Q_r, {\cal L} (q,p)\} &=& \frac{\sigma s_r}{N} \left[
-\{q_i, Q_r\}\{{\cal H}, p_i \}-\{{\cal H}, \{q_i, Q_r\}\} p_i+ \{ q_i, 
\frac{\partial q_r}{\partial t} \}p_i \right] \nonumber \\
&=& \left[ -\{{\cal H}, ~~\}+ \frac{\partial}{\partial t} \right] p_i \{ q_i, 
Q_r\} \frac{\sigma s_r}{N} \\
&=& \left[-\{{\cal H}, ~~\}+\frac{\partial}{\partial t} \right] p_i \delta_c q_i \nonumber
\eq

\bi

\noi This can be expressed in terms of Noether's conserved charge (see Eqs. 16
and 18): 

\be
-\frac{\sigma s_r}{N} \{Q_r, {\cal L} (q,p)\} = \left[-\{ {\cal H}, ~~\} +
\frac{\partial}{\partial t} \right](-{\cal L} (q,p) \delta t +\Lambda (q))
\en

\bi

The basic assumptions we will make in this section are the following (in the 
next section we will prove the validity of the two last assumptions):

\begin{itemize}
\item Only regular systems are considered.
\item In phase space, the transformation satisfies the condition 
$(q^{g_1})^{g_2} = q^{g_{12}}$ and $(p^{g_1})^{g_2} = p^{g_{12}}$.
\item The central extension $L_{rs}$ is momentum independent. (See eq. (33), 
below).
\end{itemize}

\bi

In order to show the appearance of a central extension, we begin calculating
the variation of the Lagrangian under a finite transformation. The finite 
transformation are built starting from (29), and (26). In terms of the 
intermediary variables

\bq
q^m_i (q,p) &=& exp \left[-\frac{m\sigma s_r}{N}\{Q_r,~~~\}
\right] q_i , \nonumber \\
p^m_i (q,p) &=& exp \left[-\frac{m\sigma s_r}{N} \{Q_r,~~~\}
\right] p_i , \nonumber
\eq 

\bi

\noi the variation of the Lagrangian (25) is given by:

\[ {\cal L} (q^{g(\sigma ,s)}, p^{g(\sigma ,s)}) -{\cal L} (q,p) =
\left[ -\{H,~~\}+\frac{\partial}{\partial t} \right] \sum^{N-1}_{m=1}
\Lambda^T (q^m,p^m g(\frac{\sigma}{N},s))+{\cal O} \Big(\frac{1}{N}\Big), \]

\bi

\noi where $\Lambda^T$ has been defined  by
\[\Lambda^T (q,p,g(\frac{\sigma}{N},s))=(\Lambda^r(q) -{\cal L}(q,p)\delta^r t)
s_r \sigma /N. \]

\noi We are interested in the $N\to \infty$ limit, for which the sum will
approach an integral. This is neatly seen writing $\Lambda^T =\Lambda^T_r
\delta\alpha_r$, where $\delta\alpha_r$ stands for the infinitesimal 
parameter associated to the transformations (19) and (20). Using the $\sigma ,
s$ parametrization (see discussion beneath Eq. (22)) we can write $\Lambda^T =
\Lambda^T_r s_r \frac{\sigma}{N}\,\,\overrightarrow{_{_{N\to \infty}}}\,\,\Lambda^T_r s_r d\beta$ and therefore

\bi

\[ {\cal L} (q^{g(\sigma ,s)}, p^{g(\sigma ,s)}) - {\cal L} (q,p) = \]  
\be
\left[ -\{H, ~~\}+\frac{\partial}{\partial t} \right] \int^\sigma_0
d\beta\Lambda^T_r (q^{g(\beta ,s)} p^{g (p,s)}) s_r=-\{H, \Lambda^T_f\} + 
\frac{\partial\Lambda^T_f}{\partial t} , 
\en

\bi

\noi where

\bq
\Lambda^T_f &=&\Lambda^T_f(q,p, g(\sigma ,s))= \int^\sigma_0 d\beta \Lambda^T_r
(q^{g(\beta ,s)} p^{g(\beta ,s)}) s_r . \nonumber \\
&=& \int^\sigma_0 d\beta(\Lambda_r(q^{g(\beta ,s)})- {\cal L}(q^{g(\beta ,s)},
p^{g(\beta ,s)})\delta^r t) s_r \nonumber
\eq

\bi

\noi Thus, the variation of the ``phase-space" Lagrangian turns out to be 
given by a time Lie derivative of the ``surface" term $\Lambda^T_f$.

\bi

The central point of this section relies on the observation that the
variation of the Lagrangian under finite transformations defines a
coboundary in phase space (Eq. 24 for $n=0$).

\bi

\[ \Delta {\cal L} (q,p)= {\cal L} (q^g, p^g)- {\cal L} (q,p) . \]

\noi Applying the coboundary operation to (30), and using the property
$\Delta^2=0$, we obtain:

\bq
\Delta \Delta {\cal L} (q,p) &=& 0=\Delta \left[-\{H, \Lambda^T_f\} +
\frac{\partial\Lambda^T_f}{\partial t} \right] \nonumber \\
&=&-\{ H, \Delta\Lambda^T_f\}+ \frac{\partial\Delta\Lambda^T_f}{\partial t}.
\eq

\bi

\noi The last equality can be verified by considering the explicit definition
of the coboundary operation. Eq. (31) tell us that $\omega_2 (q, p, g_1,q_2)
\equiv \Delta\Lambda^T_f (q,p,g_1)$ is conserved in time. 

\bi

The coboundary $\omega_2 (q,p, g_1,g_2)$ depends on two group elements, 
$q$ and $p$. For infinitesimal $g_1$ and $g_2$ we will parametrize the 
difference of two such coboundaries as

\be
D\omega_2 (q,p,g_1,g_2)= \omega_2 (q,p, g_1, g_2)-\omega_2 (q,p, g_2,g_1)= 
L_{rs} \alpha_{1r} \alpha_{2s} ,
\en

\bi

\noi moreover, using Eq. (24) with $n=1$  

\bq
\omega_2(q,p, g_1,g_2)-\omega_2 (q,p, g_2,g_1) &=& (\Lambda_f^T(q^{g_1}, 
p^{g_1},g_2)-\Lambda_f^T (q,q,g_2)) \nonumber \\
&+& (\Lambda^T_f(q,p,g_1)-\Lambda^T_f(q^{g_2}, p^{g_2}, g_1)) \nonumber \\
&-& \Lambda^T_f(q,p, g_{12})+ \Lambda^T_f (q,p, g_{21}). \nonumber
\eq

\bi

\noi In terms of Noether's conserved charges

\bq
-\Lambda^T_f(q,p, g_{12})+\Lambda^T_f (q,p, g_{21}) &=& Q_r\alpha_{12r} - Q_r\alpha_{21r} - p\delta_c^{g_{12}} q+ p\delta_c^{g_{21}} q , \nonumber \\
\Lambda^T_f (q^{g_1}, p^{g_1}, g_2) - \Lambda^T_f (q,p, g_2) &=&-\{\alpha_{1r} 
Q_r,\Lambda^T_f (q,p, g_2)\}=-\{\alpha_{1r} Q_r, \alpha_{2s} \Lambda^T_s\} . \nonumber
\eq

\bi

\noi Thus, we obtain:

\bq
\omega_2 (q,p, g_1,g_2)-\omega_2 (q,p, g_2,g_1) &=& -\alpha_{1r} \alpha_{2s}\{ 
Q_r,\Lambda^T_s \}+ \alpha_{2r} \alpha_{1s} \{ Q_r,\Lambda^T_s \}+ \nonumber \\
& & Q_r (\alpha_{12r}-\alpha_{21r})+p (\delta_c^{g_{21}}q-\delta_c^{g_{12}}
q). \nonumber
\eq

\bi

\noi The Poisson brackets in the expression are evaluated by expressing 
$\Lambda^T_s$ in terms of the charges.

\[ \omega_2 (q,p, g_1,g_2)-\omega_2(q,p, g_2, g_1)=\alpha_{1r} \alpha_{2s}
\{Q_r, Q_s\} + Q_t (\alpha_{12t}- \alpha_{21t}). \]

Using Eqs. (8,9) it is not difficult to show that $(\alpha_{12}-\alpha_{21})_t
= \alpha_{1r} \alpha_{2s} C^t_{rs}$. Comparing with (32), we finally conclude

\be
\{ Q_r, Q_s\}- C^t_{rs} Q_t = L_{rs} 
\en

\bi

The conservation of $w_2$ leads to 

\[ \frac{\partial}{\partial t} D\omega_2 +\{D \omega_2,H\} =\frac{\partial}{\partial q_i} D\omega_2 \frac{\partial H}{\partial p_i} +\frac{\partial}{\partial t} D\omega_2 \]

\bi

Using the asumption that $D\omega_2$ is momentum independent, taking the 
derivative of this expression respect to $p_i$ we conclude that

\[ \Big( \frac{\partial}{\partial q_i} D\omega_2 (q; g_1, g_2) \Big) 
\frac{\partial^2 H}{\partial p_j \partial p_i} = 0 . \]   

\bi

\noi Since we are considering regular systems, the Hessian 
$\frac{\partial^2 H}{\partial p_j \partial p_i}$ has no zero modes, which 
requires 

\[ \frac{\partial}{\partial q_i} L_{rs}=0 . \]

\bi

\noi Therefore the $L_{rs}$ are coordinate and momentum independent. (From the
conservation of $\omega_2$ it also follows $\frac{\partial L_{rs}}{\partial t}
=0)$.

\bi

The movement in two dimensions of a particle in a homogeneous magnetic
field provides an example where the approach so far developed can be 
applied. The system under consideration is described by the Lagrangian:

\[ {\cal L} (q,\dot q)=\frac{M}{2} \sum_i \dot q^2_i + \frac{e}{c} A_i (q)
\dot q_i, \qquad\qquad\quad i= 1,2 \]

\noi where

\[ A_i(q)=\frac{B}{2}\epsilon_{ij}q_j; \qquad\qquad\partial_i A_i=0,
\qquad\qquad {\rm and} \qquad\qquad \epsilon_{ij} = \left(\matrix{0 & -1\cr
1 & 0\cr} \right) . \]

\bi

\noi The symmetry involved  in this problem, is the translation group,
defined by the transformations:

\be
q^\prime (t) = f_i (q,t, \alpha)= q_i(t)+\alpha_i .
\en

\noi The variation of the Lagrangian under (34) is

\[ \delta {\cal L}=\frac{d}{dt} \Big(-\frac{e}{c} A_i(q) \alpha_i \Big) . \]

\bi 

\noi The $\sigma ,s$ parametrization is achieved by introducing

\[ s_i =\frac{\alpha_i}{\sqrt{\alpha_i\alpha_i}} -, \qquad \sigma=
\sqrt{\alpha_i\alpha_i} , \qquad\qquad i=1,2 , \]

\noi and

\[ q^{g(\beta ,s)} = q_i+\beta s_i \]

\bi

\noi The finite cochain is given by:

\[ \Lambda_f=\int^\sigma_0 d\beta\Lambda_i s_i=\int^\sigma_0d\beta
\Big(-\frac{e}{c} A_i (q)s_i \Big)=-\frac{e}{c} A_i(q) \alpha_i .\]

\bi

\bi

\noi Given $\Lambda_f(q,g_1)$, it is straightforward to calculate

\[\omega_2(g_1,g_2)-\omega_2(g_2,g_1)=B\epsilon_{ij}\alpha_{1i} \alpha_{2j} \]

\bi

\noi Comparing with (32), we get the central charge $L_{ij}= B\epsilon_{ij}$.

\bi

This result is easily verified. In configuration space, translations in 
orthogonal directions commute. On the other hand, in phase space, the Poisson 
brackets of Noether's charges results in $\{Q_i, Q_j\}= B\epsilon_{ij}$. 

\bi
\bi
\bi

\section{Central Charges}

\bi
\bi

In this section we present an alternative derivation of the central
extension of the algebra, which is not based on the cochain structure,
and furthermore has the advantage of showing that the central extension 
$L_{rs}$ depends only on the coordinates. 

\bi

Consider the Poisson bracket of Noether's charges (16):

\bq
\{Q_r,Q_s\}&=&\{p_i\delta^r q_i,p_j \delta^sq_j\}-\{Q_r,H\delta^s t\}-
\{H\delta^r t, Q_s\} \nonumber \\
&-& \{\Lambda_r,p_j\delta^s q_j\}-\{p_j\delta^r q_j,\Lambda_s \}, \nonumber
\eq

\noi this expression is obtained taking into account that the time variation 
is $q$-independent and therefore $\{Q_r,\delta t\}=0$. Using charge 
conservation (18) we obtain

\bq
\{Q_r,Q_s\}&=&\{p_i\delta^r q_i,p_j \delta^sq_j\}+p_i \Big(\delta^s t
\frac{\partial}{\partial t} \delta^r q_i-\delta^r t\frac{\partial}{\partial t}
\delta^s q_i \Big)\nonumber \\
&-&H \Big( \delta^st\frac{\partial}{\partial t}\delta^r t-\delta^rt
\frac{\partial}{\partial t}\delta^s t\Big)-\{\Lambda_r, p_j\delta^s
q_j\}-\{p_j \delta^r q_j,\Lambda_s\}. \nonumber
\eq

\bi

This result can be written in terms of the structure constants introduced in
(10a).

\be
\{Q_r, Q_s\}=C^u_{rs} Q_u + L_{rs}. 
\en

\noi where

\[ L_{rs}=\{\Lambda_s,p_i \delta^r q_i\}-\{\Lambda_r,p_i\delta^sq_i\}-
C^u_{rs}\Lambda_u \]

\bi

Notice that $L_{rs}$ will not depend on the momenta and that, as it
should be, it is antisymmetric in the $r-s$ indices. Explicit evaluation of 
the Poisson bracket taking into account that $\delta^r q$ are $p$ 
independent leads to:

\be
L_{rs} = \Big(\frac{\partial\Lambda_s}{\partial q_j}\Big)
\delta^r q_j -\Big(\frac{\partial\Lambda_r}{\partial q_j}\Big) 
\delta^s q_j -C^u_{rs}\Lambda_u. 
\en

\bi

In fact, if Noether charge is conserved, then (35) implies that $L_{rs}$
is also conserved. Indeed, the time Lie derivative of $L_{rs}$ is given
by:

\bq
-\{H,L_{rs}\}+\frac{\partial L_{rs}}{\partial t} =&-&\{H,\{Q_r, Q_s\}\} +
\frac{\partial}{\partial t} \{Q_r, Q_s\} \nonumber \\
&+&C^u_{rs}\left[ \{H,Q_u \}-\frac{\partial Q_u}{\partial t}\right] .\nonumber
\eq

\bi

\noi The use of Jacobi's identity and charge conservation, simplifies this 
expression to

\[ -\{H,L_{rs}\}+\frac{\partial L_{rs}}{\partial t}=\{Q_s,\{H,Q_r\}\}+
\{Q_r,\{Q_s,H\}\}+\frac{\partial}{\partial t} \{Q_r ,Q_s\}=0. \]

\bi

\noi In order to proof the central extension character of $L_{rs}$ it will
be sufficient to show that $L_{rs}$ is $q$ and $p$ independent. (From Eq. (36)
is, already clear that $L_{rs}$ is $p$ independent). To this end consider the 
time Lie derivative of $L_{rs}$.

\[ 0=\{L_{rs},H\}+\frac{\partial L_{rs}}{\partial t} =\frac{\partial H}{\partial p_j}\,\, \frac{\partial L_{rs}}{\partial q_j} +\frac{\partial L_{rs}}{\partial t}. \]

\noi Taking the derivative of this expression respect $p_i$ we obtain:

\[ \frac{\partial^2H}{\partial p_i\partial p_j}\,\,\frac{\partial L_{rs}}{\partial q_j}= 0 , \]

\bi

\noi since we restraint our analysis to regular systems, the Hessian 
can not have zero modes, which implies

\[\frac{\partial L_{RS}}{\partial q_j}=0. \]

\bi

Thus, we have shown that the Poisson bracket of Noether's charges can acquires 
only coordinate and momentum independent central extensions. This result 
justify the second assumption of the previous section. In fact, the third 
assumption can also be validated. To this end consider the difference of two 
successive transformations applied in reserved order:

\bq
\{Q_r,\{Q_s,\,\,\}\}&-&\{Q_s,\{Q_r, \,\,\,\}\} \nonumber \\
=\{\{Q_r,Q_s\},\,\,\} &=& \{C_{rs}^t Q_t +L^{rs}, \,\, \}\nonumber \\
&=& \{C_{rs}^t Q_t,\,\, \} \nonumber
\eq

\noi The last equality follows from the $q$ and $p$ independence of the 
$L_{rs}$ central charges. Thus the central charges have no effect whatsoever  
on the analogous of the Baker-Campbell-Hausdorf formula, therefore 
$(q^{g_1})^{g_2} = q^{g_{12}}$ and $(p^{g_1})^{g_2} = p^{g_{12}}$.

\bi

As an application of this approach, let us consider a free particle and 
the Galilei symmetry group. It is well known that the mass of the particle is 
involved in the algebra of the group and it is considered as a central 
extension [8]. The system under consideration is described by:

\[{\cal L}=\frac{M}{2} \sum^3_{i=1} \dot q^2_i , \qquad\qquad {\cal H}=
\sum^3_{i=1} \frac{p^2_i}{2M} . \]

\bi

The Galilei transformations, lead to the infinitesimal variations

\[ \delta q_j= (\delta v_j)t+\delta a_j , \qquad\qquad \delta \dot q_j= 
\delta v_j .\] 

\bi

\noi The $\delta q_j$ must be considered as the combination of two independent 
variations. A pure boost characterized by the parameters $(\delta v_j)$ and 
pure translations $(\delta a_j)$

\bq
\delta^r q_j (boost) &\equiv& \frac{\delta q_j}{\delta v_r} = t\delta_{jr},\nonumber \\
\delta^r q_j (trans) &\equiv& \frac{\delta q_j}{\delta a_r} =\delta_{jr} .\nonumber
\eq

\bi

\noi For infinitesimal transformations, the variation of the Lagrangian is:

\[\delta{\cal L} =\frac{d}{dt} (Mq_i \delta v_i)\]

\noi Thus, in this case, $\Lambda =Mq_i\delta v_i= \Lambda^{boost}_r \delta 
v_r+ \Lambda^{trans}_r \delta a_r$. Clearly $\Lambda^{boost}_r = Mq_r$ and 
$\Lambda^{trans}_r=0$. Noether's theorem leads to the independent conserved 
charges:  

\bi

\[Q_r = p_rt-M q_r ,\qquad\qquad P_r= p_r\qquad\qquad r= 1,2,3. \]

The Poisson brackets of these charges are:

\[ \{Q_r, Q_s\} =0 , \qquad\qquad \{ P_r, P_s\}=0,\qquad\qquad
\{ P_r, Q_s \} = M \delta_{rs} .\]

\bi

On the other hand, according to our discussion, the central extension -if it 
exist- should be given by (36). It is straightforward to show using (16)
that for this example $C^u_{rs}=0$. Futhermore, if the indices $r$ and
$s$ refer both to boost, or both to translations $L_{rs}=0$. So, the
only possibility left is:

\bi

\[ L_{rs}=\Big(\frac{\partial \Lambda^{boost}_s}{\partial q_j}\Big)\delta^r q_j
(trans)-\Big(\frac{\partial\Lambda^{trans}_r}{\partial q_j}\Big)
\delta^r q_j (boost) \]
\[ =M\delta_{sj}\delta_{jr} = M\delta_{rs} \]

\noi Therefore, we conclude that the mass is a central extension.

\newpage
 \begin{center} {\large \bf REFERENCES}
\end{center}
\small
\begin{itemize}
\item[1.-] R. Jackiw, Comments Nucl. Part. Phys. 15 (1985) 99-116.  
	   see also Y.S. Wu and A. Zee. Phys. Lett. B 152 (1985) 98.
\item[2.-] For a review see R. Jackiw, in {\it ``Relativity Groups and 
	   Topology II"}, eds., B.S. DeWitt and R. Stora (North-Holland, 
	   Amsterdam, 1984).
\item[3.-] A. Cabo, J.L. Lucio M., Phys. Lett. A 219 (1996) 155-161.   
\item[4.-] A. Cabo, J.L. Lucio M. and M. Napsuciale. Ann. Phys. 244 (1995)
	   1-11.
\item[5.-] A. Cabo, J.L. Lucio M. and H. Mercado,  to appear in Am. Jour. of 
	   Phys. (temptatively in January 1993).
\item[6.-] J. Govaerts {\it ``Hamiltonian Quantization and Constrained 
	   Dynamics"}, Leuven Notes in Mathematical and Theoretical Physics 
	   Vol. 4. Series B: Theoretical Physics, Leuven University Press, 
	   Belgium (1991).
\item[7.-] S. Treinman, R. Jackiw, B. Zumino and E. Witten, {\it ``Current 
	   Algebra and Anomalies"}, (Princenton University Press, New Jersey, 
	   1985).
\item[8.-] J.A. de Azcarraga, {\it `` Wess-Zumino Terms, Extended Algebras and
	   Anomalies in Classical Physics"}, proceedings of the AMS-IMS-SIAM.
	   Summer Research Conference on Mathematical Aspects of Classical
	   Field Theory, University of Washington at Seatle July 1991.
\end{itemize} 
\end{document}